# Document Selection in a Distributed Search Engine Architecture

[1]Ibrahim AlShourbaji, [2]Samaher Al-Janabi and [3]Ahmed Patel

[1]Computer Network Department, Computer Science and Information System College, Jazan University, Jazan 82822-6649, Saudi Arabia

[2]Department of Information Networks, Faculty of Information Technology,University of Babylon,
Babylon, Hilla 00964, Iraq

[3]Visiting Professor
School of Computing and Information Systems, Faculty of Science, Engineering and Computing, Kingston University,
Kingston upon Thames KT1 2EE, United Kingdom

## Abstract

Distributed Search Engine Architecture (DSEA) hosts numerous independent topic-specific search engines and selects a subset of the databases to search within the architecture. The objective of this approach is to reduce the amount of space needed to perform a search by querying only a subset of the total data available. In order to manipulate data across many databases, it is most efficient to identify a smaller subset of databases that would be most likely to return the data of specific interest that can then be examined in greater detail. The selection index has been most commonly used as a method for choosing the most applicable databases as it captures broad information about each database and its indexed documents. Employing this type of database allows the researcher to find information more quickly, not only with less cost, but it also minimizes the potential for biases. This paper investigates the effectiveness of different databases selected within the framework and scope of the distributed search engine architecture. The purpose of the study is to improve the quality of distributed information retrieval.

## 1. INTRODUCTION

Search engines on the internet are the most relevant and commonly used tools in searching, accessing and gathering information. Because of the volume of information out there, topic-specific search engines are being developed to be selective in searching and finding only the most relevant information [1]. The idea behind this is information retrieval that provides the representation, organization, storage and access to information items [2]. Organization and representation of the information items should provide the user with easy access to the information in whichthe users are interested and to have that information available to them when needed. Most importantly, to improve the quality of information retrieval, we have to improve the means that wereused for searching on the Web. Popular Search engines information retrieval on the Web include: Google, Yahoo, Altavista, Excite, Lycos and HotBot [3]. Search engineshave become an indispensable service to navigate on the web in an effective manner. Search engines contain three key parts: (1) a databank of information items, (2) an actualprogram designed to search those information items, and (3) a sequence of programs responsible for how the results will be presented [4]. To use search engines, searchers submit queries as sequences of terms that describe the content they are interested in.In return, the search engine generates pages of results containing lists of potentially relevant Web contents, including hypertext links to access them. Search engines mitigate the problem of finding content among billions of pages. Currently, they can generate a set of results in under a second. The speed of search engines keep rapidly increasingwhile their scalability and efficiency are becoming more and more crucial [5]. The levels of security and privacy and attempts to limit online advertising could impactthe ability of the users to reach the Web content for which they are searching as rapidly and confidently in the near future [6, 7].

DSEA is one of the means that is designed to improve information retrieval time, efficiency, and accuracy. DSEA acts as a single search system, despite the fact the search engines are owned and even controlled by different entities [8]. Their ability to set up a resource-rich system, by aggregating resources, will increase their advantage over the search engine systems. Furthermore, there are three basic activities or tasks that need to be undertaken with DSEA. First, an appropriate collection of databases need to be identified as the most applicable to search.. Next, the selected database is searched. Finally, the results are amalgamated into a single, organized response. While each of the three tasks is important and considered in this research, the primary focus will be on the task of selecting a specific database to search, or "the database selection problem [8]."

In short, this research will examine the process of providing a query to an applicable database in order to obtain a useful and organized ranking of results. Core of the DSEA is how the query is forwarded and processed in collaboration with multiple search engines to ensure that efficient processing and high-quality results. The passively discovered content is ranked with respect to a query, taking into account both the popularity and the aging of the content [9]. The main purpose of this study is to minimize the number of search engines accessed in order to reduce the huge spawning of searching and disorders for all search engines.This requires sending queries to anonly a sub-set of search engines with a target topic or keywords in a query request. In the DSEA, every search engine learns from the others accessed and learns from and ranks the effectiveness of the results for each query. With this approach, each query is selectively directed to a specific search engine(s) depending on the potential for an accurate and quick return of information.

In this paper, the overall architecture of the DSEA is discussed, based on the topic specific search engines, which is the core part of DSEA. At a higher level, the search engine functionality is spread across multiple data centers, with the main goal of meeting the performance expectations of users. The remainder of this paper is structured as follows: The next section provides related work to this study. The research method will then be detailed, followed by a brief introduction of the framework and architecture of the DSEA system. The methods and algorithms will be explained, followed by the experimental design, implementation, and its results. The paper concludes with a discussion as well as the future directions of this research.

## 2. RELATED WORK

COntent Aware Searching retrieval and sTreaming (COAST) project aims at deploying a content-centric network on top of the existing network infrastructure. This enables better use of network resources, by accessing internet content without referring to a particular copy of the data. This new architecture also enables novel features for search. The COAST search engine follows a distributed approach: Several search sites are deployed in various geographical locations and pair wise communicates to provide a search service collaboratively. This model from current search engine architecture, in which each site is independent, provides search engine functionalities by itself. In the COAST project, the researchers present the design of each component of the COAST search engine, namely, the crawler, an indexer, and the query processor. They first present an interaction between the distributed search sites. In particular, they show how the COAST search engine benefits valuable information from the COAST overlay better to identify the interests and needs of end users. Finally, the authors propose metrics associated with experiments that enable them to evaluate the COAST search engine architecture and assess its efficiency and effectiveness as well [10].

## 3. RESEARCH METHODOLOGY

In order to enhance the accuracy of information retrieval on the Web, in this study a customs method is proposed that relies on the surrounding text of a URL. To achieve this objective, CORI net is utilized for selecting the most suitable databases and returning the list of services, ranked by how well they match the user's requests [11]. In the light of this, the design and implementation of this system is a prototype.From the results of this experiment, it will be affirmed that the framework, architecture, design, and implementation of this prototype are effective and can satisfy the main objective of this study.

## 4. CONCEPTUAL FRAMEWORK

Figure 1 shows the overall conceptual framework of the DSEA system. The development of the Distributed Search Architecture system for the Topic-Specific Web Search has three main components, which work together to provide search services: The search brokers, the service directory and the topic-specific search engines. A key part of the system is the search broker. The main goal of the search broker is to offer a simple access point for users to communicate with the proposed system. It accepts search queries from the users and identifies the best search engine for that query that is being chosen from the broker's list of search engines in the system. The service directory of a topic-specific search engine interactswith the search broker for the specified search term and determinesthe most appropriate, cost-effective, search engines for the specific request of the individual user based on text provided in the query. These low-cost search engines can deliveruseful information about a specific query from each user.Seamlessly, the search broker sends the search engines to the designated topic-specific search engine, resulting in an up to the moment, accurate indexing of the Web on a specific subject. Using indexing, required search engines can be found bythe client aggregating search results from multiple databases and storing them in a new temporary database to display the results to the user on a Web browser user interface [11].

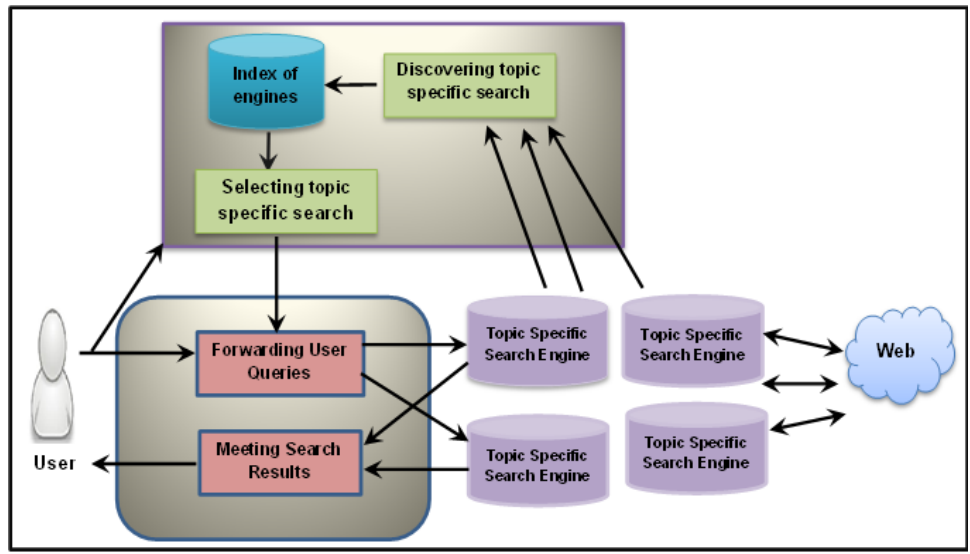

**Fig. 1.** Proposed DSEA conceptual framework and its system's components

## 4.1 DESCRIPTION OF THE SYSTEM'S COMPONENTS

**User** is anindividualwho has an inquiry and/or needs to locate documents using the DSEA system.

**Broker (Search for documents):** Figure2 shows an overview of the search scenarios. The document search scenario proceeds as follows:

1. The user (searcher) connects to one of the system's brokers with a Web browser and submits a request. This request contains the description of the desired document and other parameters (such as *Resource Description*, *Time to Live*, *Number of Results*, *Service Quality*, *etc.*) that influence the processing of the request.
2. The broker sends the user's request to a service directory, asking which databases can best serve this request. If the target database is already known (e.g., specified by the user), then:
    a) The directory returns a list of databases.
    b) The broker forwards the user's request to the selected database.
    c) The database returns the results to the broker. These results consist of document URLs and a set of keywords from the documents containing the user-specified search keywords.
3. The broker collects the results and returns them to the user via the Web interface.

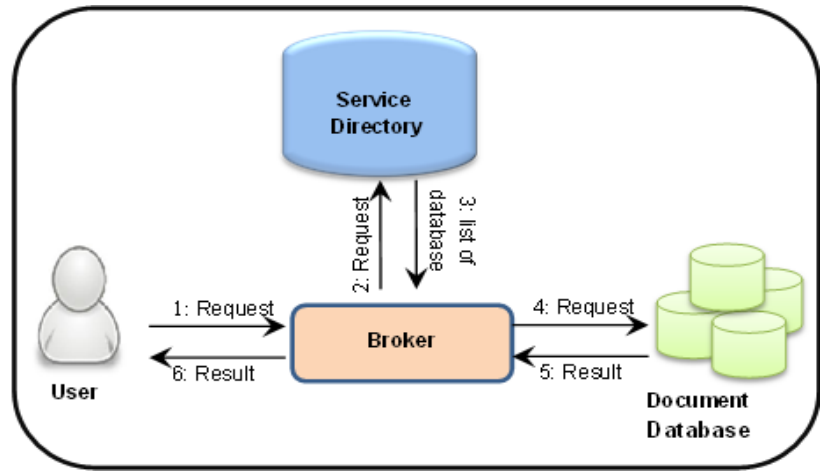

**Fig. 2.** An overview of search scenarios of a document

**SERVICE DIRECTORY**

The service directory provides the functionality necessary for searching document services and selecting the best ones for particular user queries. This increases the opportunities for the user to get relevant results while at the same time decreasing the network traffic and the load on the document database servers. The Service Directory also takes into account constraints on the time, number of results, price, and quality. For the user, it appears as a stand-alone search engine in terms of both style and appearance of the interface. In fact, the directory receives queries in nearly the same way that a typical document database might. Except, that the directory also provides a ranked list of the database names and URLs instead of a list of documents and URLs. The design of the service directory includes the directory searcher and the directory index module [12]:

- The directory searcher is responsible for processing service search requests, selecting the most suitable databases, and returning the list of services, ranked by how well they match the user's request.The algorithms used in this module is CORI net [11].
- The directory index was used for storage of the service descriptions and providing the necessary information to the directory searcher.

MySQL is used by the pilot system's service directory implementation to store the index service descriptions of the searched databases. In fact, MySQL is commonly used across a board range of Web applications given its response rates within [12].

**TOPIC SPECIFIC SEARCH ENGINES (DOCUMENT DATABASE)**

Document databases are central to the DSEA operation as it is responsible for the familiar search engine appearance to the user looking to access specific information.The efficiency is improved by the use of a Web bot the document database can use to identify and store more popular topic specific documents, URLs, and rankings from the Web and/or local data.

The "per-attribute word indexes" in the document store includes the documents having been identified as having the requested keywords. In addition, the results of the search request are sorted by relevance a displayed for the user.

## 5. METHODS AND ALGORITHMS

This section briefly introduces the methods and algorithms, which are wide spread approaches for databases selection and Information Retrieval.

### 5.1 PORTER STEMMING ALGORITHM

Information Retrieval (IR) often involves automatically removing suffixes [13]. In an IR setting, a selection of documents is individually described by words included in the title, the abstract, and throughout the document. Each document can be, in essence, identified by the presence of words or terms within, disregarding suffixes. Terms including a common root typically convey the same gist, for example:

> CONNECT
> CONNECTED
> CONNECTING
> CONNECTION
> CONNECTIONS

Frequently, the performance of an IR system improves if terms are semantically grouped by conflating them into a *single term* to enhance the searching operation. Different suffixes -ED, -ING, ION, IONS, for example, can be removed from the root word. In the case of the example above, each term would be reduced to CONNECT. This process will also reduce the total count of search terms in the system, and therefore, the size and complexity of the data within the system were likewise reduced.Considering this, Porter's algorithm [15] is primarily essential for two reasons: A) , The algorithm allows the search term to be conflated in a way that is practical for a variety of languages. B) The concept of stemming, itself, has recently been recognized as a potential area of interest, beyond being just a part of a larger system

### 5.2 CORI net ALGORITHM

CORI net is a prevalent approach for database selection. It is one of the most popular benefits estimators often used for database selection strategies. CORI net is defined as a collection ranking algorithms for the inquiry retrieval system. It uses an interface network to rank collection.

Currently, parallel IR systems are required to manage separate directories to provide the level of results the user is seeking. In order to do this cost- effectively, a query broker module could be developed that would route a query to a specific document collection relative to the topic being queried. This document selection technique would, necessarily, be more efficient overall as it would be searching only a subset of all available documents. Such a selection technique involves ranking, then selecting the best database relative to each user's request considering the specific query posed. In order to be successful, the database selection algorithm is a pivotal part of the service directory. The database selection function is run when the user search request is submitted to the service directory. The request includes data used by the database selection algorithm: Desired content description, constraints on time and price, desired number of results (documents or hits), and desired number of databases.

Selection databases are executed in two steps:

1. The CORI network algorithm will be used to identify and rank the information contained in all of the databases registered in the service directory according to the requested topic description.
2. A utility value is used to rend the databases according to calculations of the rest of the parameters: Time, number of results and price.

**Using the following statistic to rank the collections. Figure 3 shows the input information for the CORI net algorithm:**

- Set $T$ of term $t_i$, where $i = 1, \ldots, n$.
- Set $C$ of database $db_j$, where $j = 1, \ldots, m$.
- Matrix $(n \times m)$ of document frequencies $df_{ij}$, if term $t_i$ occurs in $k$ documents in database $db_j$ then $df_{ij}$ otherwise $df_{ij} = 0$.
- The belief $p(t_i|db_j)$ in database $db_j$ due to observing term $t_i$, which is determined by:
  $df\_ij = d\_t + (1 - d\_t).\log(\llbracket df \rrbracket \_ij + 0.5)/\log(\llbracket df \rrbracket \_j^{\wedge}max + 1.0)$
-

$$icf_j = \log((|C| + 0.5)/cf_i)/\log(|C| + 1.0)$$

$$p(t_i|db_j) = d_b + (1 - d_b).\hat{d}f_{ij}.icf_j$$

Where $df_j^{max}$ is the document frequency of the most frequent term in $db_j$, $\hat{d}f_{ij}$ is the weighted document frequency of the term $t_i$ in $db_j$, $|C|$ is the number of databases, $cf_i$ is the collection.

1. initialise $d_t = 0.4$ and $d_b = 0.4$
2. **for all** term $t_i \in T$ **do**
3. Set $cf_i$ to the collection frequency of the term $t_i$ (i.e. the number of databases, in which this term occurs)
4. Calculate inverse collection frequency
   $$icf_i = \log og|C| + 0.5)/cf_i)/\log ogC| + 1.0)$$
5. **end for**
6. **for all** database $db_i \in C$ **do**
7. **for all** term $t_i \in T$ **do**
8. calculate weighted document frequency
   $$\hat{d}f_{ij} = d_t + (1 - d_t).\log ogdf_{ij} + 0.5)/\log ogdf_j^{max} + 1.0)$$
9. calculate belief $p(t_i|db_j)$ in database $db_j$ due to observing term $t_i$
   $$p(t_i|db_j) = d_b + (1 - d_b).\hat{d}f_{ij}.icf_j$$
10. store the value of $p(t_i|db_j)$
11. end for
12. end for

Fig. 3.CORI net Algorithm

## 6. EXPERIMENTAL DESIGN

The experimental design has four phases that are related to each other as shown in Figure 4. The first one is shown after the user enters a keyword and then the administrator collects the datasets, creates a number of databases, and then compute the frequency for each database. The second phase is utilizing the CORI net algorithm for selecting the best database from a number of databases after creating the service-directory, which is joined with the broker. The third phase is a normal case of the user to request his keyword(s) and wait for the results. The last phase,the stemming process stems the keywords, and then, connects with a specific database from the service-directory. Finally, we select *max ranking* instance keywords from the service-directoryto display the results of the user.

The broker helps the users to find an appropriate search engines for any given search requests by providing the system with a Web-based user interface. The broker communicates with the service directory and search engines to satisfy the user's requests according to the user's needs. The broker then submits the user's requests to a selected directory, retrieves a list of best matching search engines and re-submits the request to one or more of the search engines depending on its frequency. The results returned from the search engine(s) are then passed on to the user. The broker, therefore, performs the requested brokerage and query propagation by using the information from the service directory.Figure 5 below shows the experimental design for the DSEA system.

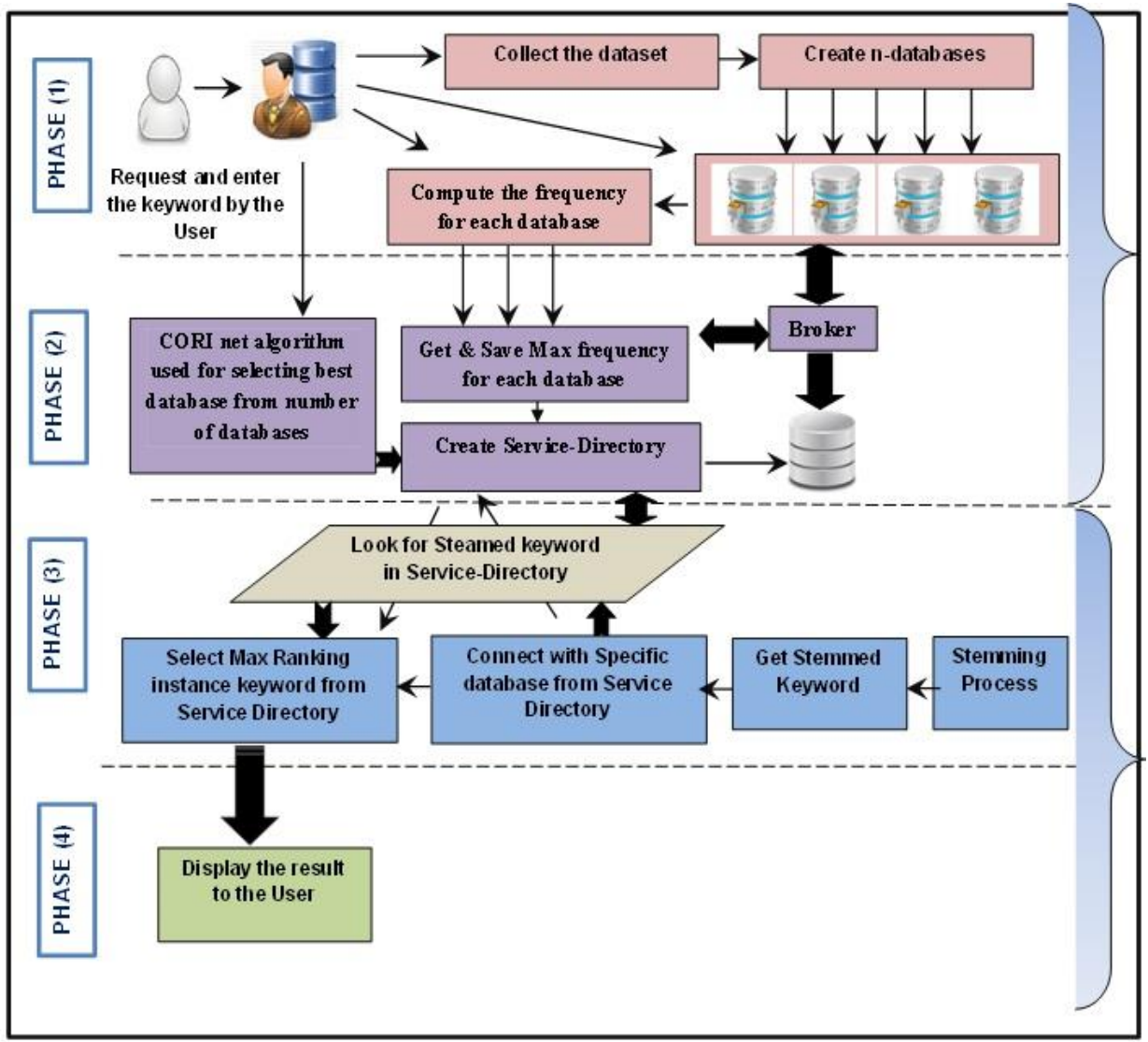

**Fig. 4.** Experimental Design Process

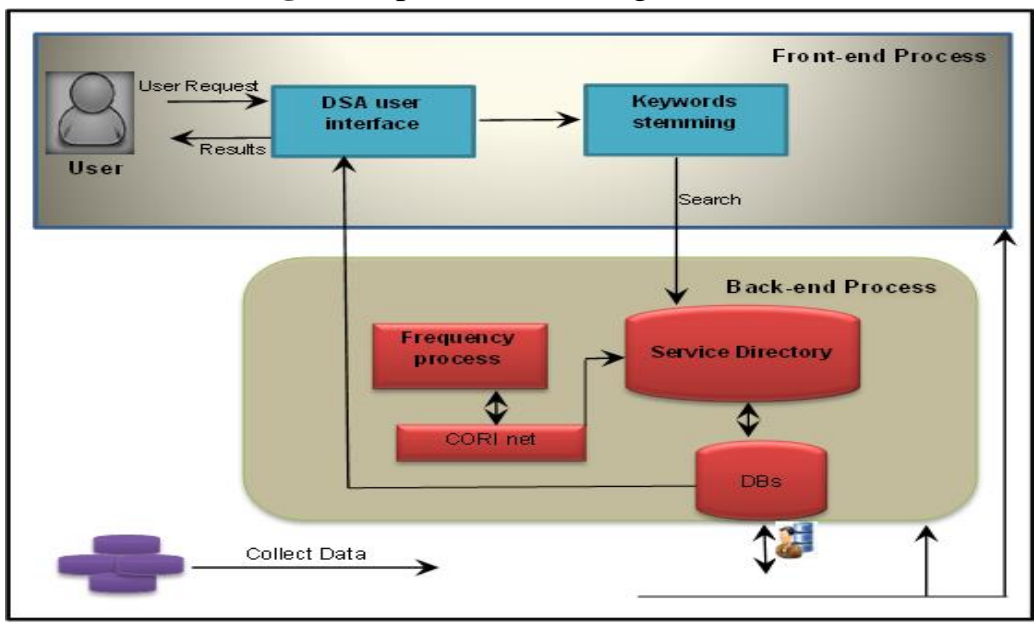

**Fig. 5.** Experimental Design

## 7. IMPLEMENTATION

The experimental design executes according to the following phases:

### 7.1 SERVICE DIRECTORY

Our service directory system provides many functions to select the best documents in our databases which deal with the user query. The following are the main requirements in our system:

- Look for the documents: This component is used to select, rank and list the best documents in our directories based on a user's request.
- Clear the previous sessions: Prepare the current searching activity by removing the previous activities to avoid searching conflicts.
- Discover reports: This component is to discover if there are any well ranked documents found in our directories and report the searching activities in order to allow us to enhance our contents.
- Evaluation management: This component records and parameterizes the available documents to provide accuracy and speed in our searching process.

The function configure () implements the Configuration of the service directory. It also manages and retrieves the frequencies of the keywords. As explain in the following pseudo code.

```
Public void configure() throws SQLException , IOException
{
Vector v=new Vector ();
String[] db_name=DataReader.getData();
for (int i = 0; i<db_name.length; i++)
  {
statement.executeUpdate("use "+db_name[i]);
rs=statement.executeQuery("select count(*), keyword  from data group by keyword");
 While (rs.next ()) do
   {    DataBean data = new DataBean();
  data.setDB_name(db_name[i]);
  data.setTerm_name(rs.getString(2));
  data.setFrequancy(rs.getString(1)) ;// continue
```

```
v.add(data);// continue
}
}
statement.executeUpdate("use service directory");
for (int i1=0;i1<v.size();i1++)
  {begin for
DataBean  data =  (DataBean)v.get(i1);
 Insert (" insert into db_data_temp (db_name, term_name, frequency) values ('"+data.getDB_ name () +"', '"+ data. getTerm_ name () +"', '"+data. getFrequency()+"')");
    }* end for
statement.executeUpdate("insert    intodb_    data    (db_name, term_name,    frequency)    select    db_name,term_name,max (frequency) from db_data_temp group by term_name");
}* end function
```

## 7.2   STEMMING COMPONENT

This code implements the Porter Stemming algorithm [13], which receives the keywords from the user and stems them [14]. We then run our enhanced algorithm that joins with a service directory to search all keywords, which contains the stemmed keywords. This technique also helps in retrieving the results for the user. The below pseudo code of the class Porter algorithm that performsthe Stemming process.

```
public class Porter
{
 public static void main (String[] args)
 {
        String str = "education";
        str=str.toLowerCase();
        str = stripAffixes( String str );
        str = stripPrefixes(str);
        if (str != "" )
str = stripSuffixes(str);
```

```
        String suffixes = { "al", "ance", "ence", "er", "ic", "able",
"ible", "ant", "ement", "ment", "ent", "sion", "tion","ou", "ism",
"ate", "iti", "ous", "ive", "ize", "ise"};

};
NewString stem = new NewString();
for ( int index = 0 ; index<suffixes.length; index++ )
{
if ( hasSuffix ( str, suffixes[index], stem ) )
{
if ( measure ( stem.str ) > 1 )
 {
str = stem.str;
returnstr;
    }
}                                      // continue
}                                      // continue
 System.out.print("the   output   is        =>>           " +
stripAffixes("education"));
  }
}
```

## 7.3 DATA COLLECTION

The DSEA system was implemented in Java Server Pages JSP, where all the sub-systems are programmed as classes. In terms of object-oriented concepts, the system has several classes, each of which corresponds to one of the main components in the system. However, the document databases relatedtothe search engines are managed by MySQL.We conduct the experiments with the five databases created. As an example,*Education*can beused as a topic to be queried in a topic-oriented system. In this example, a specific database would be responsible for maintaining each collection, from which a "training set" would be taken for the topic, Education.Considering this example, the number of topics with matching collections are compiled in Table 1.

**Table 1.** Topic-specific experimental databases

| N | Name of database | Description | Record Count |
|---|---|---|---|
| 1 | $DB_1$ | Number of URLs in $DB_1$ | 119 |
| 2 | $DB_2$ | Number of URLs in $DB_2$ | 105 |
| 3 | $DB_3$ | Number of URLs in $DB_3$ | 81 |
| 4 | $DB_4$ | Number of URLs in $DB_4$ | 108 |
| 5 | $DB_5$ | Number of URLs in $DB_5$ | 125 |

## 7.4 RESULTS

The system was developed, it was run successfully on the provided data set and performed as planned. The output of the system was the result of a search by the topic related keywords and their URLs, as well as the priority score associated with each keyword. This score conveyed the frequency of the keyword. The results are shown in Figure 6 as a response to a user query.

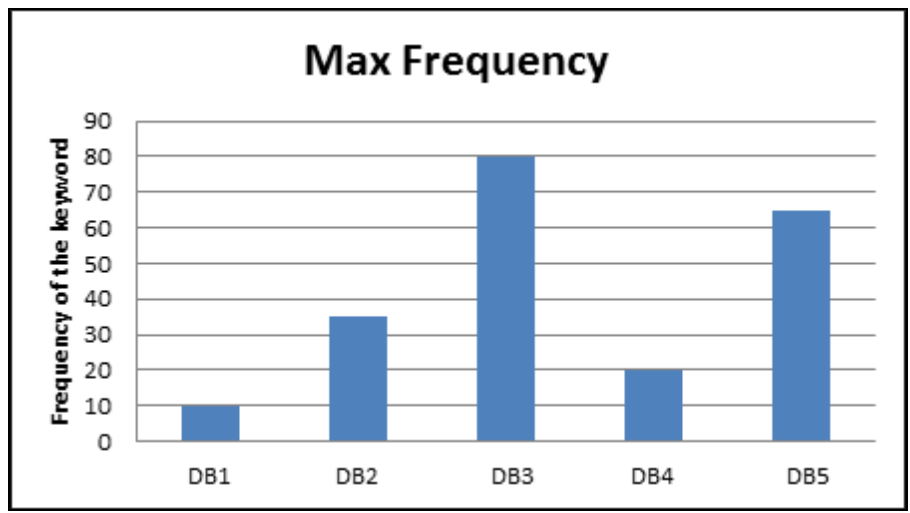

**Fig. 6.** The frequency of a keyword for all databases in the system

## 8. DISCUSSION

DSEA is a novel approach providesusers with the information they want quickly and cost-effectively. By utilizing DSEA, the user can overcome the potential obstacles that could result from the exponential increase the number of online searches and the number of search engines available to use for Web searches. Prior to use DSEA, the best search outcomes, based on studies of distributed search architecture, were provided by a Service Directory (SD). Service directories, at that time, maintained mapping between SEs and the resources available on the Web. The growthof distributed search architecture concertedon the service directory. This allowed the cost to be minimizedby limiting the needs for bandwidth and computation power by utilizingother existing technologies like the meta-search engines. The components of the system were intendedto be consistent with each other and to provide the user with the information sought. The system was populated with databases including documents about one topic: Query processing, or analyzing a query and comparing it to indexes to find the relevant items. The user entered a keyword into the DSA interface. The broker was responsible for query processing in response to receiving a keyword entered by the user.  The indexed web pages were then scanned for the keywords and evaluated.  The SD performed the evaluation to determinethe rank, and ultimately, order of display of results to the user based on that level. The SD used the CORI-net algorithm to analyze and rank databases according to relevance by counting the number of times the requested word or phrase was identified in the indices. The results of testing showed an increase in effectiveness by applying the algorithm. The bandwidth cost was reduced, while the users acquired information and data that was more accurate and relevant as the CORI net algorithm was used with a smaller scope. It was tasked  only to evaluate a specified set of documents that already contained known keywords and were already ranked in the SD. The broker was able to  select efficiently the most appropriate databases considering the query and the maximum frequency of the keyword(s). The designed model required A) The development of the DSEA using an SD to guide searches to the SEs with the applicable topic-specific documents.B) building a system that could accommodate a larger number of users searching the massive amount of data available on the World Wide Web (WWW) concurrently. Sorting the responses in more focused topic-specific SEs  allowed for both efficient and effective searching. This achieved the most acceptable performance and was consistent with the efficiency and quality requirements of the users. The future research into the advancement of Search Engines (SEs) would be beneficial as an additional need has been identified to develop methods to scale in two orthogonal dimensions: Data and access. Any solution advanced in the future must be able to the mind-boggling billions of search engines housing millions if notbillions of documents and be able to accommodate millions of accesses a day.

# CONCLUSION

The service directory of a topic-specific search engine interacts with the search broker for the specified search term and determines the most appropriate, cost-effective, search engines for the specific request of the individual user based on text provided in the query. These low-cost search engines can deliveruseful information about a specific query from each user.Seamlessly, the search broker sends the search engines to the designated topic-specific search engine, resulting in an up to the moment, accurate indexing of the Web on a specific subject.

# REFERENCE

1. Bajat. M. K. B., 2011. Evaluation of the Proposed Topic Specific Search Engine in Geostatistics. Ostrava, 1:23 – 26.

2. Sanderson. M. and W. B. Croft, 2012. The history of information retrieval research. Proceedings of the IEEE, Special Centennial Issue, 100:1444-1451.

3. Chernov., S., P. Serdyukov, M. Bender, S. Michel, G. Weikum and C. Zimmer, 2007. Database selection and result merging in p2p web search. Databases, Information Systems, and Peer-to-Peer Computing. Springer Berlin Heidelberg, 4125: 26-37.

4. Silvestri. F., 2010. Mining query logs: Turning search usage data into knowledge, Foundations and Trends in Information Retrieval, 4: 1-174.

5. Cambazoglu, B. B. and R. Baeza-Yates, 2011. Scalability challenges in web search engines. In Advanced topics in information retrieval, 33: 27-50.

6. Patel, A., AlShourbaji, I. and Al-Janabi, S, 2014. Enhance Business Promotion for Enterprises with Mashup Technology. Middle East Journal of Scientific Research, 22: 291-299.

7. Patel, A., Al-Janabi, S., AlShourbaji, I. and Pedersen, J, 2015. A novel methodology towards a trusted environment in mashup web applications. Computers & Security, 49: 107-122.

8. Cambazoglu. B. B., V Plachouras. and R. Baeza-Yates, 2009. Quantifying performance and quality gains in distributed web search engines. In Proceedings of the 32nd international ACM SIGIR conference on Research and development in information retrieval, 411-418.

9. Jonassen .S., 2013. Efficient Query Processing in Distributed Search Engines. Thesis for the degree of Philosophiae Doctor at Norwegian University of Science and Technology.

10. Bai. X., T. Morel., I. K. Zahariadis. and G. M. Arumaithurai, 2012. D3. 4: Distributed Searching in Future Internet. Project No: FP7- ICT-248036-COAST.

11. Callan. J. P., Z. Lu. and W. B. Croft, 1995. Searching distributed collections with inference networks. In Proceedings of the 18th annual international ACM SIGIR conference on Research and development in information retrieval, 21-28.

12. Hofstede. R., 2009. Performance measurements of NfDump and MySQL and development of a SURFmap plug-in for NfSen. Bachlor assignement, University of Twente.

13. Porter M. F., 1980. An algorithm for suffix stripping. Program: electronic library and information systems, 14(3): 130-137.

14. Sogrine M. and A. Patel, 2003. Evaluating database selection algorithms for distributed search. In Proceedings of the 2003 ACM symposium on Applied computing, 817-822.

15. Willett. P., 2006. The Porter stemming algorithm: then and now. Program: electronic library and information systems, 40: 219-223.